# Modélisation analytique et numérique de structures sandwich de type carton ondulé

S. Allaoui[*]   Z. Aboura[**], M.L. Benzeggagh[*], N. Talbi[***] & R. Ayad[***]

[*] Université de Technologie de Compiègne. Laboratoire de Roberval de Mécanique
CNRS-UPRES A 6066 B.P 20529 F- 60205 Compiègne Cedex. samir.allaoui@utc.fr
[**] IUT de Tremblay en France Paris 8 - LAHP/ERBEM Rue de la Râperie 93290 Tremblay-en-France
[***] ESIEC Esp. Rolland Garros BP 1029 51686 Reims Cedex 2

**Résumé :**

*Le carton ondulé est un sandwich utilisé le plus souvent pour l'emballage sous forme de caisse. Sa recyclabilité et sa biodégradation font que c'est l'un des emballages les plus utilisé. Sa mise en œuvre sous forme de structure sandwich rend les études numériques du produit (caisses) très lourdes. Des modèles analytiques et numériques ont été développés pour prédire le comportement mécanique du matériau homogénéisé. Suite à cela les études deviennent plus souples et moins coûteuses. Un modèle analytique d'homogénéisation basé sur la théorie de stratification a été proposé dans de précédentes études. Ce dernier permet de prédire le comportement élastique du carton ondulé. Ce travail propose, dans un premier temps, une extension de ce modèle, au cas de carton ondulé composé de plus de trois constituants. Une simulation numérique de flexion permettant d'inclure l'effet de cisaillement est proposée. Par la suite le modèle analytique est étendu pour prédire le comportement anélastique du carton ondulé.*

**Abstract :**

*Corrugated cardboard is a sandwich used generally for packing in the form of case. It's the most used material for packing because it can be recycled as well as biodegradable. It's implementation in form of structure, makes the numerical studies of the product (cases) very heavy. Analytical and numerical models were developed to predict the mechanical behavior of the homogenized material. Following that the studies become more simple and less expensive. An analytical model of homogenisation based on the classical laminate theory has been proposed in a previous studies. It predicts the elastic behavior of the corrugated cardboard. This work purposes an extension of this model. Initially, It will be applied at the case of corrugated cardboard composed of more than three constituants. After that, a numerical study of bending witch take account of the effect of shear is proposed. Finally, the anlytical model will be used to predict the elastic and anelastic behavior of the corrugated cardboard.*

**Mots clefs :** carton ondulé – modélisation – homogénéisation – sandwich

# 1 Introduction

Le carton ondulé, essentiellement utilisé pour l'emballage, est un matériau complexe composé de plus de deux feuilles. Les feuilles planes extérieures sont appelées couvertures ou peaux et les feuilles cannelées appelées cannelures ou âmes. Le carton ondulé est mis en œuvre sous forme de structure sandwich (matériau) avant de réaliser le produit final (caisse d'emballage). Le problème de l'analyse du comportement mécanique de ce matériau peut être abordé dans un premier temps, en considérant le carton ondulé en tant que structure et en le caractérisant par des normes s'apparentant aux normes relatives aux structures sandwich composites ou métalliques. Cette démarche mène à un maillage total de la structure (peaux et cannelure) afin de modéliser le comportement de l'emballage dans son ensemble. Cette approche rend les études onéreuses et lourdes en temps de calcul. Une seconde approche, proposée par Aboura et al [1], considère le sandwich carton ondulé comme un matériau monolithique. Cette démarche nécessite le développement de modèle analytique ou numérique [2] permettant d'obtenir les paramètres homogénéisés élastiques du matériau et ainsi de simplifier les approches numériques(*figure 1*).

Dans ce travail on propose d'étendre le modèle analytique proposé par Aboura et al [1] dans un premier temps au cas des cartons ondulé composés de plus de trois feuilles. Par la suite ce modèle sera développé pour prédire le comportement inélastique du matériau. De plus une étude numérique utilisant un nouvel élément coque DMTS [3,4] sera réalisée pour simuler le comportement en flexion du carton ondulé homogénéisé en incluant l'effet du cisaillement. Les résultats seront confrontés à des résultats expérimentaux.

# 2 Modèle analytique

Le modèle de l'étude est inspiré d'autres travaux sur homogénéisation des matériaux composites à renfort tissé [5, 6]. Ces travaux prennent en compte l'effet de l'ondulation et la continuité des mèches pour la détermination des propriétés élastiques des matériaux composites. Le principe du modèle proposé repose sur l'application de la théorie de stratification à chaque partie infinitésimale de la structure à homogénéiser. Dans le cas du sandwich carton ondulé, le volume élémentaire représentatif (VER) présenté dans la *figure 2* est discrétisé en micro-éléments. Pour chaque élément *dx*, les matrices de membrane, de couplage et de flexion locales (A, B, et D) sont déterminées. Les propriétés élastiques globales sont ensuite obtenues après homogénéisation de l'ensemble.

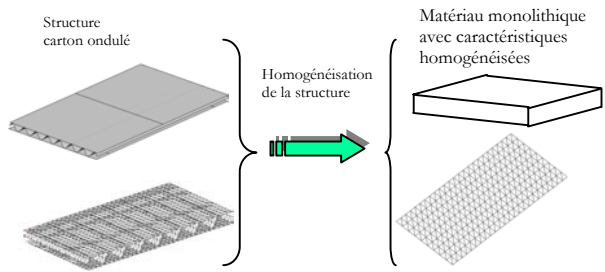
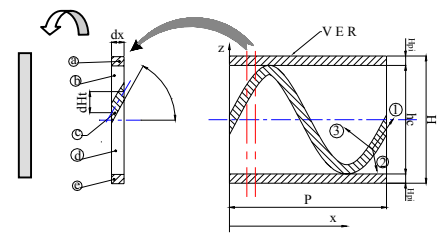

FIG. 1 – Approche d'homogénéisation                    FIG. 2 – Volume Elémentaire Représentatif

Ce modèle a été utilisé, pour la modélisation du comportement élastique d'un carton ondulé double face (DF, deux peaux et une cannelure) et a donné satisfaction. Dans cette étude il va être étendu pour le cas d'un carton ondulé double-double face (DD, trois peaux et deux cannelures). Deux voies sont possibles. Soit la structure est considérée dans son ensemble et ensuite discrétisée comme précédemment afin d'extraire les matrices (A, B, et D), soit de considérer le double-double face comme étant une stratification de 2 doubles faces. Dans ce cas la démarche consiste à homogénéiser chaque strate individuellement puis à appliquer la théorie de la stratification à l'ensemble. Cette deuxième démarche est intéressante dans le sens où la prédiction du comportement de sandwichs plus complexes tel que le triple face ou un hybride (superposition de différents double face) est facilitée puisqu'elle ne nécessite pas la réécriture des équations d'homogénéisation.

Le carton ondulé utilisé pour la modélisation est symétrique et composé de trois peaux et de deux ondulations dont les caractéristiques mécaniques et géométriques sont illustrées dans le *tableau1*. L'épaisseur totale du sandwich est de 8,032 mm. A noter que $E_{MD}$ correspond au module de Young dans le sens machine (sens d'ondulation) et $E_{CD}$ à celui dans le sens transverse.

Les résultats du modèle montrent que les deux configurations d'homogénéisation sont en bonne corrélation. Hormis le module de Young dans le sens transverse ($E_{CD}$) où on note une différence de 8 Mpa, correspondant à une erreur de 2,5% ce qui est minime, les autres paramètres sont identiques(*tableau 2*). Le modèle peut donc être appliqué sur des cartons ondulés de plus de trois constituants en procédant soit par homogénéisation directe ou indirecte.

|  | Epaisseur (mm) | Emd (Mpa) | Ecd (Mpa) | $\nu_{xy}$ | $\nu_{yx}$ |
|---|---|---|---|---|---|
| Peau | 0,264 | 4514,53 | 1895,83 | 0,282 | 0,215 |
| ondulation | 0,21 | 4703,75 | 1854,3 | 0,358 | 0,088 |

Tab1 – Données géométriques et mécaniques des composants du carton ondulé

|  | Emd (Mpa) | Ecd (Mpa) | $\nu_{xy}$ | $G_{xy(MD)}$ |
|---|---|---|---|---|
| Homogénéisation par couche | 607,1 | 354,87 | 0,24208 | 172,87 |
| Homogénéisation direct | 607,1 | 345,9 | 0,24208 | 172,87 |

Tab2 – Résultats du modèle analytique

## 3 Modèle élément fini

L'idée d'une modélisation 3D d'un produit en carton ondulé (caisse) serait trop onéreuse tant sur le plan éléments que CPU. L'approche serait donc de modéliser ce type de structure par des éléments coque après homogénéisation. L'élément utilisé est un élément de coque, le DMTS à trois nœuds et six degrés de liberté par nœud. Ce dernier est obtenu par une superposition d'un élément de membrane (CST) et d'un élément de plaque épais (DMTP). L'élément de plaque basé sur un modèle en déplacement DDMT en l'état, proposé par Katili [7] et qui a été obtenu en utilisant une procédure d'équivalence avec le modèle mixte modifié MiSP3+/M[5][6], ne peut être utilisé pour modéliser des plaques orthotropes mono ou multicouches. Il a été nécessaire d'intervenir au niveau du bord élémentaire pour modifier les hypothèses discrètes de Mindlin, utilisées pour éliminer localement les rotations $\Delta\beta_{sk}$. qui sont initialement définies aux nœuds milieu des cotés (*figure 4*).

Afin de tester la validité de la démarche, des essais de flexion trois points à différents élancements L/h (rapport entre la longueur entre appuis et l'épaisseur) allant de 10 à 50 ont été simulés sur des plaques de carton ondulé double-faces. Le comportement du matériau obtenu lors de ces essais a été comparé au comportement obtenu dans les mêmes conditions lors des tests expérimentaux. Ces derniers ont été réalisés dans le sens CD. Les éprouvettes utilisées ont une largeur de 60 mm et une épaisseur de 4.1 mm et dont les caractéristiques des peaux sont illustrées sur le *tableau 3*. Les paramètres mécaniques des plaques utilisés pour l'étude numériques sont les paramètres homogénéisés obtenus par le modèle analytique (*tableau 3*). La comparaison des deux comportements s'est faite sur la base de la pente (Δ) des courbes charge-déplacement de flexion.

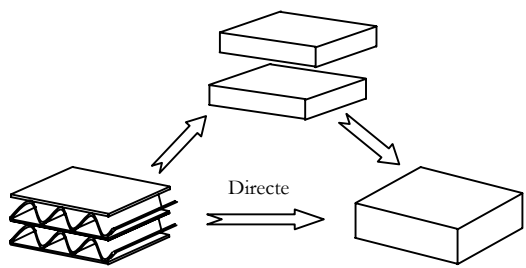

FIG. 3 – Homogénéisation d'un carton ondulé double-double face

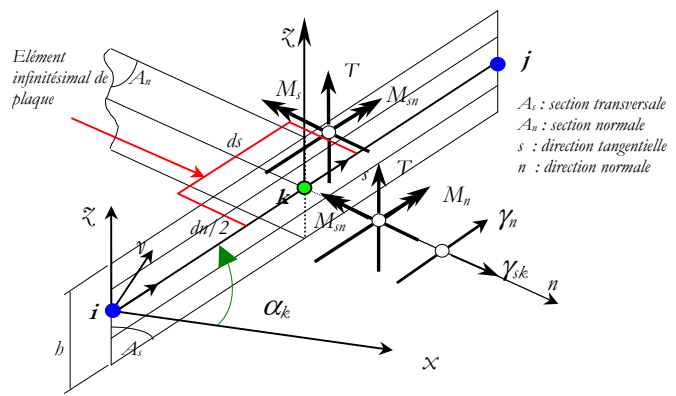

FIG. 4 – Hypothèses de Mindlin sur le côté ij d'un élément fini multicouche

Le *Tableau 4* montre que le comportement du carton ondulé est bien reproduit une erreur maximale entre les pentes expérimentales($\Delta_{EXP}$) et les pentes numériques($\Delta_{EF}$) de 17%.

Cet écart peut être dû à l'écart entre les caractéristiques homogénéisés et les caractéristiques expérimentales. En effet le modèle analytique présenté offre une précision de l'ordre de 10% sachant que la dispersion naturelle des résultats expérimentaux est aussi de l'ordre de 10%.

Néanmoins ce qu'il faut retenir c'est que l'effet du cisaillement introduit par des éprouvettes à faible élancement (L/h=10) est bien pris en compte par l'élément développé puisque l'écart entre l'expérience et la simulation est de 6%.

|  | $h$ | $E_{(MD)}$ | $E_{(CD)}$ | $\nu_{xy}$ | $\nu_{yx}$ | $G_{xy(MD)}$ |
|---|---|---|---|---|---|---|
| Peau supérieure | 0,235 | 4514,53 | 1895,83 | 0,282 | 0,215 | / |
| Peau ondulée | 0,19 | 4458,6 | 1944,5 | 0,277 | 0,115 | / |
| Peau inférieure | 0,235 | 4703,75 | 1854,3 | 0,358 | 0,088 | / |
| Carton (homogénéisées) | 4,1 | 717,7467 | 389,969 | 0,24 | 0,1304 | 202,709 |

Tab 3 – Caractéristiques des peaux et du carton après homogénéisation.

| L/h | 10 | 20 | 30 | 40 | 50 |
|---|---|---|---|---|---|
| $\Delta_{EXP}=P/W_{EXP}$ | 111 | 22,798 | 7,308 | 3,172 | 1,632 |
| $\Delta_{EF}=P/W_{EF}$ | 117,451 | 20,145 | 6,165 | 2,641 | 1,363 |
| Erreur % | 6 | 12 | 16 | 17 | 17 |

Tab 4 – Pente en fonction de l'élancement.

## 4   Modélisation du comportement anélastique

La modélisation analytique du comportement du carton ondulé a été réalisée dans le domaine élastique. Au-delà de cette partie élastique, le matériau a un comportement viscoélastique. L'extension du modèle analytique à la simulation du comportement au-delà de l'élasticité nécessite la connaissance des lois de comportement des constituants du carton ondulé. Dans une première approche, le modèle sera alimenté par les résultats expérimentaux issus des tests effectués sur les peaux et la cannelure [8][9].

L'application d'un chargement $(N_t)_{globale}$ sur le sandwich génère une déformation$(\varepsilon_t)_{globale}$. En prenant l'hypothèse d'isodéformation, et connaissant les rigidités locales, il est possible de remonter au tenseur des contraintes locales $(\sigma)_{locales}$ des constituants en utilisant leurs lois de comportement. Disposant des tenseurs de contraintes et de déformations, la rigidité locale est alors déterminée pour chaque niveau de chargement. Le comportement global est obtenu en procédant à chaque pas de chargement à une homogénéisation.

Un carton ondulé double face (DF) a été utilisé afin de valider le modèle. Les courbes ( σ=fonction(ε) ), obtenues en traction uni-axiale du matériau homogénéisé, dans les deux sens (CD et MD), ont été générées avec les résultats du modèle et comparées avec les courbes expérimentales de traction du sandwich. *La figure 5 (a)* montre une bonne corrélation entre le comportement simulé dans le sens MD et les résultats expérimentaux. Par contre la corrélation est moins bonne dans le sens CD. En effet après une phase élastique approchée convenablement, le modèle surestime la phase anélastique. Ce phénomène est du à l'effet de l'interface entre les deux peaux et l'ondulation, qui intervient lors de la sollicitation, et qui n'est pas pris en compte par le modèle. Cette interface est composée d'amidon servant de colle entre les sommets de l'ondulation et les peaux. Lors de l'essai de traction, la sollicitation de l'interface provoque un endommagement localisé du à la rigidité élevée de l'amidon. Ce phénomène n'intervient pas dans le sens MD car l'ondulation n'est très peu sollicitée et la réponse du sandwich est gouvernée par les peaux.

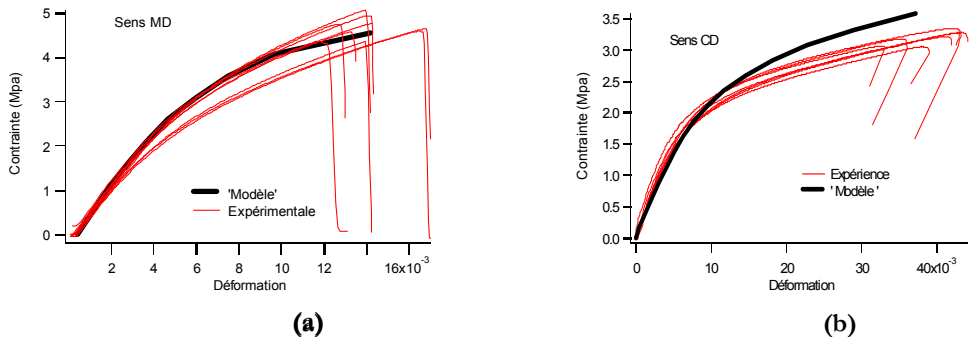

FIG. 5 – Corrélation modèle–expérience du comportement du matériau

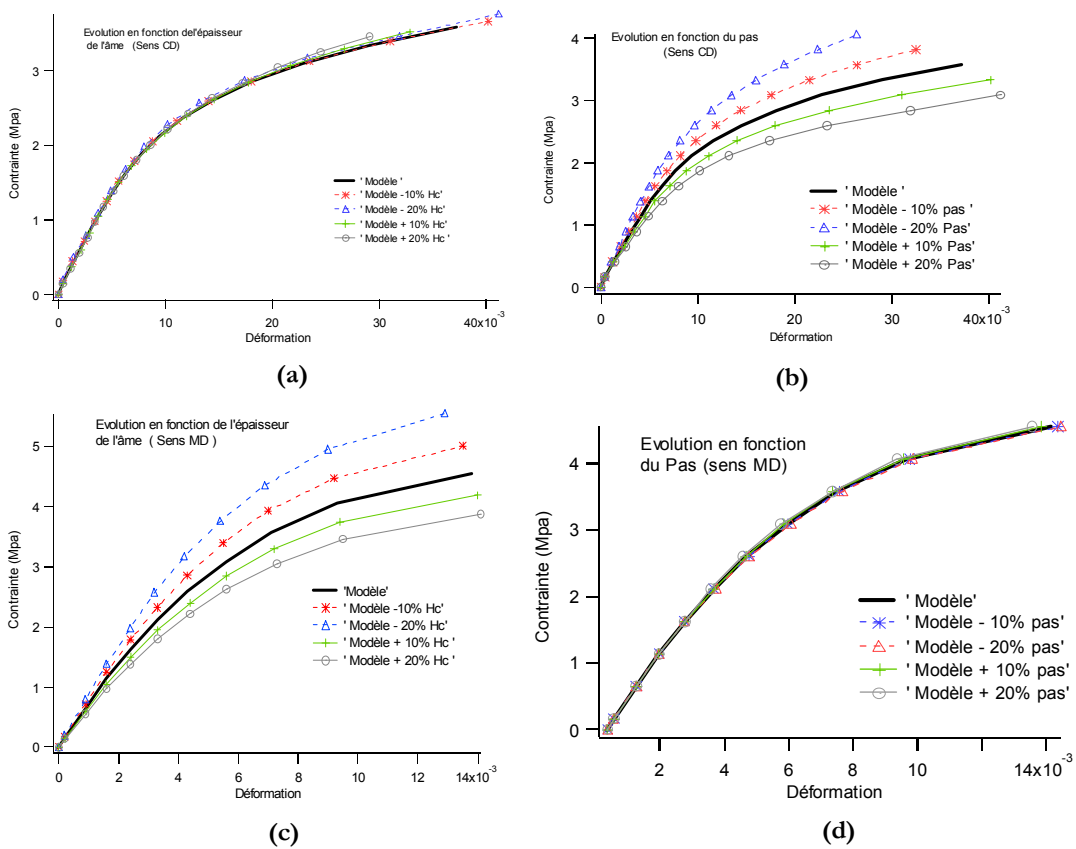

FIG. 6 – Evolution des résultats du modèle en fonction du pas de l'ondulation et de l'épaisseur de l'âme

Après la validation du modèle, une étude de l'effet des caractéristiques géométriques du carton ondulé sur le comportement global en traction uni-axiale a été effectuée.

Les paramètres étudiés sont le pas d'ondulation et l'épaisseur totale de l'âme (Hc). Cette variation est de ±10% et ±20%. Les *figures 6 (a) et (b)* montrent que le comportement du matériau dans le sens CD est plus sensible à l'évolution du pas d'ondulation qu'à l'épaisseur de l'âme. Dans le sens MD, seule l'épaisseur de l'âme influence le comportement du matériau ( *figures 6 (c) et (d)*).

Ainsi, en baissant l'épaisseur, et par conséquent en tendant vers une configuration dite micro-cannelure, la limite d'élasticité du matériau augmente ainsi que la contrainte ultime. Cet effet est plus important dans le sens MD. Ce constat est également applicable à l'effet du pas d'ondulation dans le cas d'un chargement dans le sens CD. Par contre, compte tenu du fait que la cannelure est très peu sollicitée dans le sens MD, il n'y a pratiquement aucun effet du pas d'ondulation dans ce sens de chargement.

La contrainte à rupture est inversement proportionnelle à la variation des deux paramètres alors que la déformation maxi à rupture en est proportionnelle.

## 5 Conclusion

Cette étude, qui s'inscrit dans une série de travaux consacrés à l'étude du comportement mécanique des structures sandwich de type carton ondulé, propose une extension du précédent modèle analytique aux cartons ondulés de plus de trois constituants. Elle montre que les résultats d'une superposition de couches homogénéisées sont identiques aux résultats issus d'une démarche globale considérant l'ensemble des constituants. De plus ce travail montre la validité, dans le domaine élastique, d'un nouvel élément fini le DMTS pour la prise en compte des effets de cisaillement dans le cas de ces matériaux.

L'étude s'est poursuivi par l'extension du modèle à la prédiction du comportement anélastique du carton ondulé. Les résultats obtenus approchent correctement le comportement expérimental en traction uni-axiale. Le modèle développé nécessite d'être alimenté par les lois de comportement des constituants. Dans un premier temps le comportement expérimental a été utilisé. Les travaux futurs s'orientent vers l'élaboration de lois de comportement macroscopiques des constituants.

Ces travaux permettent de proposer des études paramétriques sur l'effet des paramètres structuraux sur le comportement global de la structure dans le but d'optimiser le choix de la solution matériau.

## Réferences